\documentclass[letter]{aa}

\usepackage{graphicx}
\usepackage{txfonts}
\usepackage{natbib}
\usepackage{ulem}
\usepackage[flushleft]{threeparttable}
\usepackage{multirow}
\usepackage{gensymb}
\usepackage{scrextend}   
\usepackage[colorlinks=true, linkcolor=blue, citecolor=blue, draft=false]{hyperref}
\usepackage{xcolor}
\usepackage{CJK}
\usepackage[
singlelinecheck=false 
]{caption}%
\defcitealias{Shi2016}{S16}  

\begin{document} 

\begin{CJK*}{UTF8}{gkai}
   \title{Extremely weak CO emission in IZw 18}

\author{Luwenjia Zhou \inst{\ref{c1},\ref{c2}}\thanks{E-mail: \texttt{wenjia@nju.edu.cn}}
\and Yong Shi\inst{\ref{c1},\ref{c2}}\thanks{E-mail: \texttt{yshipku@gmail.com}}
\and Zhi-Yu Zhang\inst{\ref{c1},\ref{c2}}
\and Junzhi Wang\inst{\ref{c3}} }

\institute{School of Astronomy and Space Science, Nanjing University, Nanjing 210093, China \label{c1}
\and Key Laboratory of Modern Astronomy and Astrophysics (Nanjing University), Ministry of Education, Nanjing 210093, China \label{c2}
\and Shanghai Astronomical Observatory, Chinese Academy of Sciences, 80 Nandan Road, Shanghai 200030, Peopleʼs Republic of China\label{c3}
}

\date{Received --; accepted --}

 
  \abstract{Local metal-poor galaxies are ideal analogues of primordial galaxies with the interstellar medium (ISM)   barely being enriched with metals. However, it is unclear whether carbon monoxide remains a good tracer and coolant of molecular gas at low metallicity. Based on the observation with the upgraded Northern Extended Millimeter Array (NOEMA), we report a marginal detection of CO $J$=2-1 emission in IZw18,  pushing  the detection limit down to $L^\prime_{\rm CO(2-1)}$\,=\,3.99\,$\times$\,10$^3$\,K\,km\,s$^{-1}$\,pc$^{-2}$, which is at least 40 times lower than previous studies. As one of the most metal-poor galaxies, IZw18 shows extremely low CO content despite its vigorous star formation activity. 
  Such low CO content relative to its infrared luminosity, star formation rate, and [C\,\textsc{ii}] luminosity, compared with other galaxies,  indicates a significant change in the ISM properties at a few percent of the Solar metallicity. In particular, the high [C\,\textsc{ii}] luminosity relative to CO implies a larger molecular reservoir than the CO emitter in IZw18.
We also obtain an upper limit of the 1.3\,mm continuum, which excludes a sub-millimetre excess in IZw18.}

   \keywords{galaxies -- dwarfs:  ISM -- molecules
               }

   \maketitle
%

\section{Introduction}
Star formation occurs in molecular gas dominated by H$_2$, except perhaps for the  in the early universe. However, H$_2$ can only be excited at a temperature above 100\,K, hence it cannot be observed directly in the cold molecular gas that fuels star formation. The second most abundant molecule, CO, has been demonstrated to be a powerful tracer of  molecular clouds in galaxies, but this application becomes complicated in metal-poor environments \citep[see][for a review]{Bolatto2013}.

In the early Universe, the first galaxies formed in the primordial gas with few elements being heavier than hydrogen. Observationally, at $z$\,$\gtrsim$\,5, normal star-forming galaxies with star formation rates (SFRs) of ten to a few hundred solar masses per year indeed have a low dust content \citep{Walter2012, Capak2015}, comparable to local metal-poor galaxies. Nevertheless, the predicted CO flux in these high-$z$ galaxies is of the order of $\mu$Jy and remains challenging due to the state-of-the-art sub-millimetre array, Atacama Large Millimeter/submillimeter Array (ALMA). A detailed study on the metal-poor interstellar medium (ISM) relies on its local analogues.

Despite the extensive search for CO emissions in local dwarf galaxies \citep{Leroy2007, Schruba2012, Cormier2014, Hunt2015, Warren2015, Shi2015}, the detection rate decreases sharply in galaxies with a metallicity lower than one-fifth the Solar metallicity ($Z_\odot$)\footnote{Here we adopt Solar metallicity as  12\,+\,log(O/H)\,=\,8.7 \citep{Asplund2009}.}.  The extremely faint CO emission brings about questions as to the existence of molecular gas in such galaxies. On the other hand, CO may not be an ideal tracer of molecular gas \citep{Grenier2005, Wolfire2010, Shi2016} in low-metallicity environments. 
The interstellar radiation penetrates deeper into molecular clouds due to the low dust content. 
At such depths within the clouds, CO is more easily destroyed than H$_2$ through dissociation, and thus it ceases to be a reliable tracer.
In addition, the reduced opacity in the stellar atmosphere leads to harder radiation fields, and hence reinforces the destruction of CO molecules into atomic or singly ionised carbon.

Low CO content is then expected in metal-poor galaxies. On the other hand, exposed to the hard radiation field, CO molecules can be easily photodissociated and produce ionised carbon, [C\,\textsc{ii}]. Furthermore, [C\,\textsc{ii}] could come from ionised gas as well as neutral gas and the surface of photodissociation regions (PDRs). However, the fractions of [C\,\textsc{ii}] emission from ionised gas derived from observation constraints are systematically lower than the one from simulations \citep{Accurso2017, Cormier2019}. Therefore it is still  unclear whether it is feasible to use [C\,\textsc{ii}] to trace H$_2$ gas.   

Substantial  efforts have been made to search for the molecular gas in metal-poor galaxies \citep{Rubio2015, Oey2017, Schruba2017, Elmegreen2018}; nevertheless, among the galaxies below 10\%\,$Z_\odot$, only Sextans\,B (7\%\,$Z_\odot$) has a robust CO detection \citep{Shi2016, Shi2020}.
The impact of low metal abundance on  star formation requires further exploration of the most metal-poor galaxies. In this work, we used the upgraded NOEMA interferometer to observe the CO $J$=2-1 emission in IZw18, which is one of the galaxies with the lowest metallicities in the local Universe \citep[$\sim$\,3\%\,$Z_\odot$, ][]{Izotov1999}.
IZw18 is a proto-type of blue compact dwarfs located at a distance of 18.2\,Mpc \citep{Aloisi2007}. The active star formation therein \citep{Hunt2005} indicates the presence of molecular gas. \citet{Leroy2007} obtained an upper limit of CO(1-0) emission ($L^\prime_{\rm CO}$\,<\,10$^5$\,K\,km\,s$^{-1}$\,pc$^2$) using the Plateau de Bure Interferometer (PdBI), which however is not deep enough to conclude whether IZw 18 has a normal CO content relative to other physical properties. Here we put further constraints on the CO content in IZw18 by pushing down the detection limit of $L^\prime_{\rm CO}$ by 40 times using the Northern Extended Millimeter Array (NOEMA) after its Phase II upgrade.
\begin{table*}[ht]
\centering
\caption{Observation information.}
\begin{tabular}{l c c c c c c c c c}
\hline
\noalign{\vskip 3pt}
& \# of Antennas & \multicolumn{2}{c}{Calibrators} &  PWV$^{1}$ & $T_{\rm sys}$  & \multicolumn{2}{c}{Resulting sensitivity}\\ 
\cline{3-4} 
\cline{7-8} 
\noalign{\vskip 1pt}
&                &   Phase/Amplitude & Bandpass & &  &Line$^{2}$ & Continuum$^{3}$\\ 
\hline
\noalign{\vskip 3pt}
obs-17 & 9  & 0954+556, 1030+611 & 3C273 & 3\,mm & $\sim$156.1\,K & 1.5\,mJy/beam& $28.9\,\mu$Jy/beam\\ %
\noalign{\vskip 3pt}
obs-19 & 9  & 0925+504, 0954+556 & 3C84  & 3\,mm & $\sim$178.1\,K & 1.8\,mJy/beam& $34.5\,\mu$Jy/beam\\ 
merged &   &  &   &  &  & 1.3\,mJy/beam& $26.2\,\mu$Jy/beam\\ 
\hline
\end{tabular}
\caption*{\scriptsize 
$^{1}$Precipitable water vapour. \\
$^{2}$ For the 3D cubes smoothed to 3\,km\,s$^{-1}$ and reconstructed with natural weighting.  \\
$^{3}$ For the images tapered to a beam size of 2\arcsec.}
\label{table:obs}
\end{table*}

\begin{table}[!t]
\caption{Basic information on IZw18 used in this paper.}
\begin{tabular}{l c c c c c c c c c}
\hline
\hline
Parameter & Value & Reference\\
\hline
$V_{\rm hel}$ & 751 km\,s$^{-1}$ & (1)\\
$D$      & 18.2\,Mpc     & (2)\\
12+log(O/H)   & 7.17 &(3)\\ 
$M_{\star}$   & 4.33\,$\times$\,10$^{6}$\,M$_{\sun}$ & This paper.$^{(a)}$\\
\textit{SFR}       & 0.12\,$\pm$\,0.01\,M$_{\sun}$\,yr$^{-1}$ & This paper.$^{(b)}$\\
$S_{\rm CO(2-1)}$ & 2.48\,mJy & This paper.$^{(c)}$\\
$F_{\rm CO(2-1)}$ & 19.42\,mJy\,km\,s$^{-1}$ & This paper.$^{(d)}$\\
$L^{\prime}_{\rm CO(2-1)}$ & 3.99$\times10^3$ K\,km\,s$^{-1}$pc$^2$ &This paper.\\
$L_{\rm CO(1-0)}$ & 0.18\,L$_\sun$ & This paper.$^{(e)}$\\
$F_{\rm 1.3\,mm}$ & 181\,$\pm$\,51\,$\mu$Jy & This paper.$^{(f)}$\\
\hline
\end{tabular}
\caption*{\scriptsize
(1) NASA/IPAC Extragalaxtic Database (NED).\\
(2) \citet{Aloisi2007}.\\
(3) Main body in the south-west \citep{Izotov1999}.\\
(a) Derived from the SED fitting in Section~\ref{sec:sed}.\\
(b) Based on the emission at FUV and 24\,$\mu$m. \\
(c) Peak flux density of CO(2-1) as described in Section~\ref{sec:co21}.\\
(d) Velocity integrated line flux of CO(2-1) from 40\,km\,s$^{-1}$ to 50\,km\,s$^{-1}$.\\
(e) CO(1-0) luminosity. Assuming optically thick and thermalised $L^\prime_{\rm CO(1-0)}$\,=\,$L^\prime_{\rm CO(2-1)}$.\\
(f) 1.3\,mm continuum flux as described in Section~\ref{sec:cont}.
}
\label{table:para}
\end{table}

\section{Observations}
\label{sec:obs}
The observations of IZw18 were carried out with  NOEMA  on  December 24, 2017 (obs-17) and  March 23, 2019 (obs-19) for a total of 16\,h (11.6\,h on source) with configuration D.  We processed the data at IRAM/Grenoble using the \texttt{GILDAS} package. The final CO\,$J$\,=\,2-1 datacubes of obs17, obs18, and the merger of the two have beam sizes of 1\farcs83\,$\times$\,1\farcs60 (obs-17), 1\farcs71\,$\times$\,1\farcs41 (obs-18), and 1\farcs71\,$\times$\,1\farcs48 (merged), respectively, and a frequency resolution of  0.2MHz (0.26 km\,s$^{-1}$ at 230\,GHz). IZw18 was entirely covered by the 22$\arcsec$  (FWHM) primary beam. The details of the observation settings and the sensitivities of the data are listed in Table~\ref{table:obs}.

\section{Results}
\label{sec:results}
\begin{figure}[htbp]
\centering
        \includegraphics[width=\columnwidth]{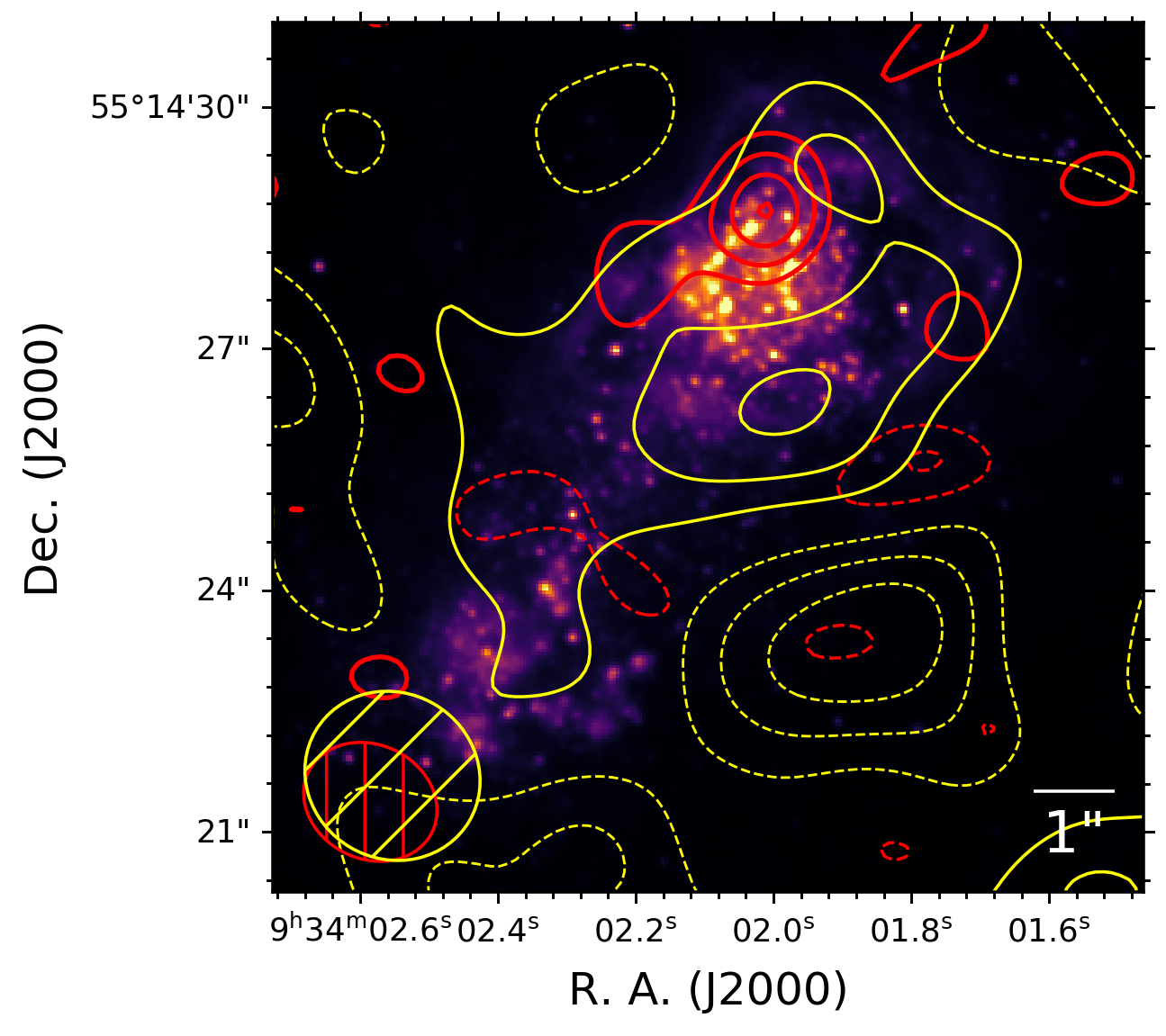} 
    \caption{ CO $J$=2-1 emission and 1.3\,mm continuum superimposed on the HST $V$-band image from \citet[][PropID: 9400]{Izotov2004}. The red contours denote the CO $J$=2-1 integrated line intensity  at the resolution of 1\farcs71\,$\times$\,1\farcs48, starting from 2$\sigma$, in increments of 1$\sigma$~(3.3\,mJy/beam) significance.  The yellow contours denote the continuum detection at 1.3\,mm at the resolution of 2\farcs21\,$\times$\,2\farcs06, starting from 1$\sigma$, in increments of 1$\sigma$ (24\,$\mu$Jy\,beam$^{-1}$) significance. Dashed contours are negative.}
\label{fig:imgV}
\end{figure}
\begin{figure}[htbp]
\centering
        \includegraphics[width=\columnwidth]{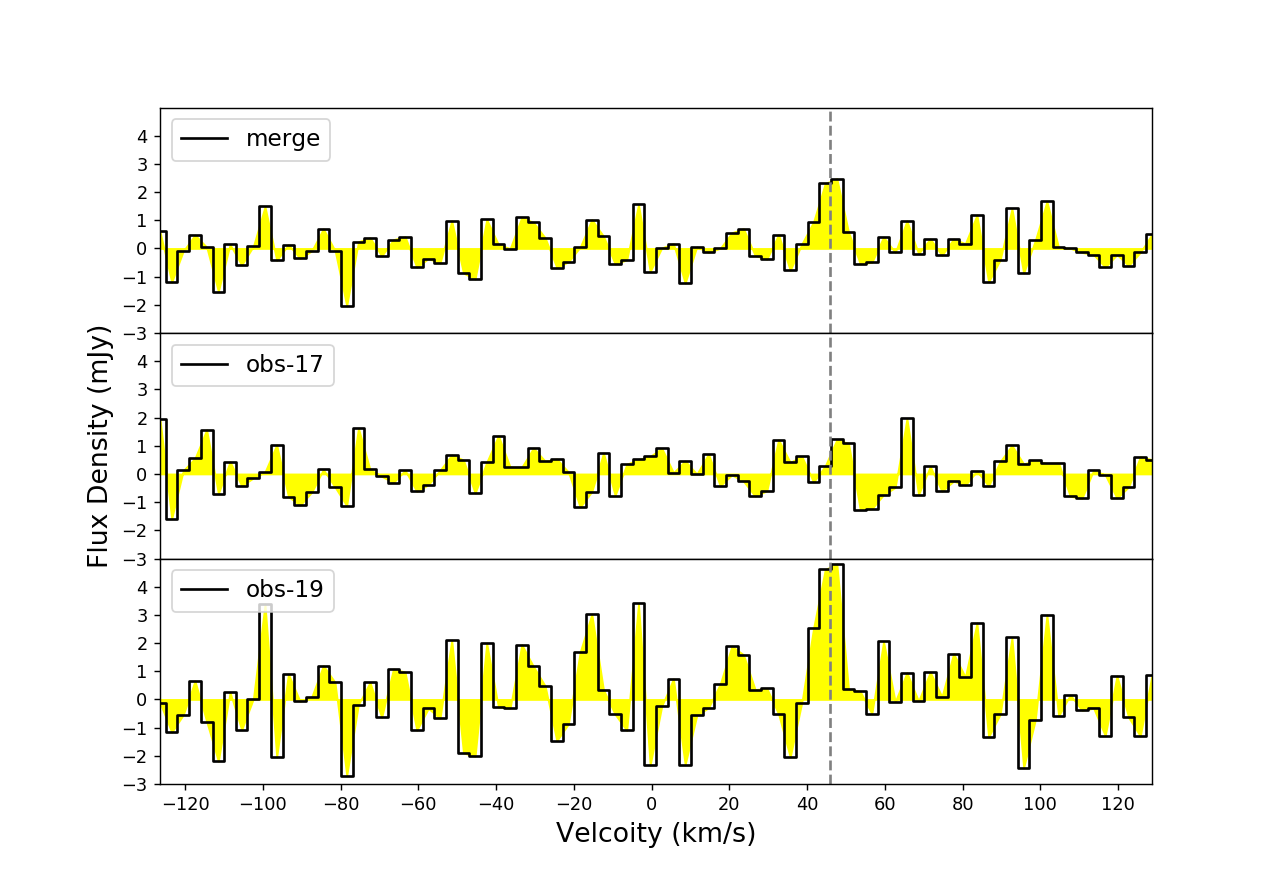} 
    \caption{CO $J$=2-1 spectra from the merged, obs-17 and obs-19 data cubes. The spectra are smoothed to a velocity resolution of 3\,km\,s$^{-1}$ and extracted from the area defined by the 3$\sigma$ contour as shown in Figure~\ref{fig:imgV}. The 1$\sigma$ uncertainty is 0.72, 0.72, and 1.39\,mJy,  respectively. The grey dashed line denotes the marginal detection at $\sim$46\,km\,s$^{-1}$.}
\label{fig:spectra}
\end{figure}

\subsection{CO $J$=2-1}
\label{sec:co21}
We smoothed the data to a velocity resolution of 3\,km\,s$^{-1}$, consistent with the CO line width found in the most metal-poor galaxies in the literature \citep{Shi2016}. We found one marginal detection ($\alpha$\,=\,9$^{\rm h}$34$^{\rm m}$02$^{\rm s}$00, $\delta$\,=\,55\degree14\arcmin28\farcs81) at the peak of the stellar emission, as indicated by the red contour \footnote{The red contours show the  CO $J$=2-1 line intensity integrated from 40\,km\,s$^{-1}$ to 50\,km\,s$^{-1}$, where the marginal detection falls, as shown in Fig.~\ref{fig:spectra}} superimposed on the Hubble Space Telescope (HST) $V$-band image (Fig.~\ref{fig:imgV})\footnote{We determined the HST astrometry in this field to be flawed by $\sim$\,0.3\arcsec, and hence negligible. }. It covers an area defined by the 3$\sigma$ contour of $\sim$1\farcs52$\,\times\,$1\farcs52, which is approximately the size of the synthesised beam. We have plotted the spectra from the merged, obs-17 and obs-19 data cubes in Fig.~\ref{fig:spectra}. A  3$\sigma$ spectral signal  falls at 46\,km\,s$^{-1}$.  The peak fluxes derived from the peak channels have signal-to-noise ratios of 3.47, 2.73, and 3.45 for the merged, obs-17 and obs-19 data cubes. We note that even though the spectrum extracted from the obs-17 data cube shows lower rms noise, the noise level in the data cube slightly deviates from a Gaussian distribution, while those from obs-19 and merged data cubes do not. Hereafter, we use the result from the merged data cube.

The CO $J$=2-1 peak flux density is $S_{\rm CO(2-1)}$\,=\,2.48\,mJy and the corresponding CO luminosity is then derived to be $L^\prime_{\rm CO(2-1)}$\,=\,3.99\,$\times$\,10$^3$\,K\,km\,s$^{-1}$\,pc$^{-2}$ using the formulation of \citet{Solomon2005}. The CO emission tends to reside in clumps  of a few parsecs at low metallicity \citep{Rubio2015,  Oey2017, Schruba2017, Shi2020}, which is much smaller than the beam size of our observation, hence we do not expect significant missing diffuse emission.

\subsection{1.3\,mm continuum}
\label{sec:cont}
We created a continuum map after removing  channels with emission. The continuum map was smoothed to have an angular resolution  of 2\farcs21\,$\times$\,2\farcs06 with an rms noise of 51\,$\mu$Jy\,beam$^{-1}$, using the \textsc{tclean task in CASA}. The  flux peak falls at $\alpha$\,=\,9$^{\rm h}$34$^{\rm m}$02$^{\rm s}$00, $\delta$\,=\,55\degree14\arcmin26\farcs30 which is between the two flux peaks of the H\,\textsc{i} emission of IZw18 \citep[][see also Appendix~\ref{sec:append}]{Lelli2012} . This is the only 3$\sigma$ signal ($F_{\rm 1.3\,mm}$\,=\,181\,$\mu$Jy) within the primary beam. Comparisons between CO $J$=2-1 emission and 1.3\,mm continuum as well as the emission  at FUV, 3.6\,$\mu$m, 24\,$\mu$m, 100\,$\mu$m, and of H\,\textsc{i} gas are shown in Appendix~\ref{sec:append}.

\section{Discussion}
\label{sec:discussion}
In this section, we discuss how  the CO luminosity and the 1.3\,mm continuum flux constrain our current understanding of the star formation process in the extremely metal-poor environment in IZw18. The physical parameters we used for IZw18 are listed in Table~\ref{table:para}.

We compare IZw18 to normal star-formimg galaxies and metal-poor galaxies in the literature. These include seven metal-poor dwarf galaxies from the $Herschel$ Dwarf Galaxy Survey \citep[DGS,][compiled in \citealt{Accurso2017}]{Cormier2014},  16 nearby dwarf galaxies  in \citet{Schruba2012} and nearby galaxies compiled therein, eight metal-poor galaxies from \citet{Hunt2015}, local galaxies with stellar mass higher than 10$^9$\,M$_\odot$ from the xCOLD GASS survey \citep{Saintonge2017}, and  massive, infrared bright, local star-forming  galaxies at Solar metallicity from \citet{Gao2004}, as well as the individual star-forming regions in  four extremely metal-poor galaxies (12\,+\,log\,(O/H)\,<\,8) from \citet{Shi2015, Shi2016}.

\subsection{SED and sub-millimetre excess}
\label{sec:sed}
We first investigated the spectral energy distribution (SED) of IZw18 taking into account the 3$\sigma$ upper limit at 1.3\,mm to put more constraints on the FIR/millimetre regime. All other photometric data were taken from \citet{Hunt2014}. We performed the SED fitting with CIGALE \citep{Boquien2019} adopting a delayed star formation history, a \citet{Draine2007} dust model, which is consistent with the method used for the star-forming regions in \citet{Shi2015,Shi2016},  and a radio component.  The infrared luminosity derived from the fit is 2.2\,$\times$\,10$^{7}$\,L$_{\odot}$, consistent with the one from \citet{Hunt2014}. The stellar mass is 4.33\,$\times$\,10$^{6}$\,M$_{\odot}$, within the uncertainty of the stellar masses derived from stellar mass-to-light ratios in the K band and 3.6\,$\mu$m \citep{Fumagalli2010, Hunt2019}, while it is around ten times lower than those derived based on the R band luminosity \citep{Lelli2012, Lelli2014}.

\begin{figure}[htbp]
\centering
        \includegraphics[width=\columnwidth]{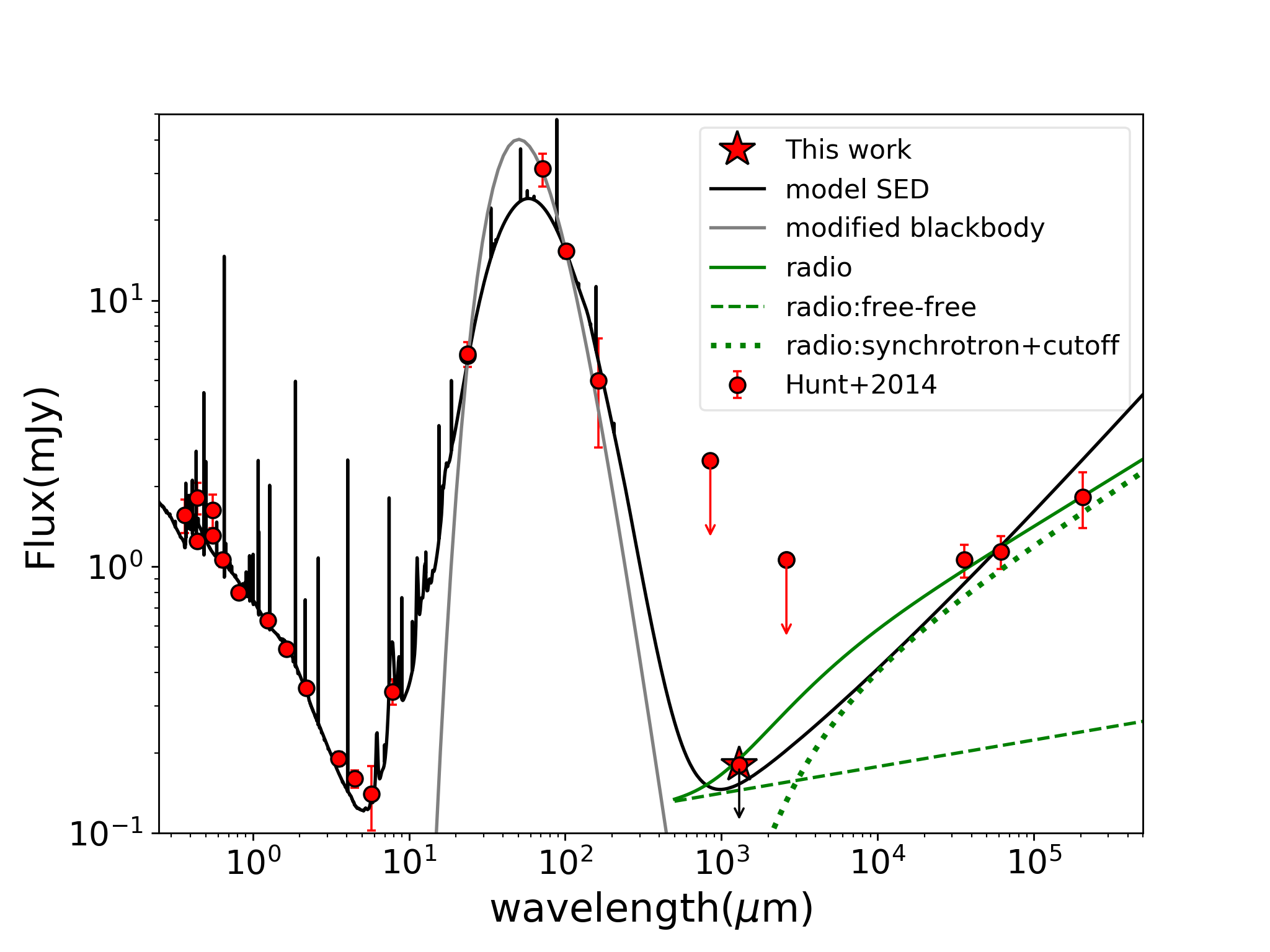} 
    \caption{SED of IZw18, with CIGALE best fit adopting a delayed star formation history, a \citet{Draine2007} dust model,  and a radio component. A modified blackbody fit is shown with a grey line. Photometric data are from \citet{Hunt2014}, except for at 1.3\,mm. The 3$\sigma$ upper limit at 1.3\,mm  is shown as in the  red star. The radio emission is also fitted with a cutoff model including a free-free component and a synchrotron component as introduced in \citet{Klein2018} (see more in the Section~\ref{sec:sed}), which is indicated by the green curves.}
\label{fig:sed}
\end{figure}

We also explored the existence of the excess at sub-millimeter wavelengths \citep[e.g.][]{Galliano2003,Lisenfeld2002,Galametz2011, Remy-Ruyer2013} in IZw18. It did not have a robust detection at the far-infrared (FIR)  wavelengths longer than 160\,$\mu$m. The 3$\sigma$ upper limit at 1.3\,mm places a strong constraint on the (sub-)millimetre end of the dust emission. We  fitted the FIR SED with the modified blackbody model and derived a lower limit of the dust emissivity $\beta$ to be 2.1, which suggests no  sub-millimetre excess in IZw18. Furthermore, taking into account the upper limit at 1.3\,mm, we notice that the radio emission shows a steeper slope ($\alpha$\,=\,-0.58), compared to the spectral indice found in \citet{Hunt2005} ($\alpha_{\rm 1.4GHz}^{\rm4.8GHz}$\,=\,-0.39, $\alpha_{\rm4.8GHz}^{\rm 8.4GHz}$\,=\,-0.13); this may be due to the high frequency cutoff of synchrotron emission from the relavitistic electrons in the star-forming regions. Therefore, we fitted the radio emission with a cutoff model as introduced in \citet{Klein2018}, by fixing the spectral index of the synchrotron emission to be $\alpha_{\rm nth}$\,=\,$\alpha_{\rm 1.4GHz}^{\rm 4.8GHz}$\,=\,-0.39 (Fig.~\ref{fig:sed}). Then the contribution of the free-free emission to the total radio emission is $f_{\rm 1.3\,mm}^{\rm ff}$\,$\sim$\,76$\%$ and $f_{\rm 2.6\,mm}^{\rm ff}$\,$\sim$\,54$\%$, higher than what was found in the literature \citep[e.g.][]{Leroy2007, Hunt2014}. We note that the cutoff frequency derived from the fit is at 145\,GHz, which is higher than the typical value of $\sim$10\,GHz found in \citet{Klein2018}.

\subsection{$L_{\rm IR}$ and SFR versus $L^\prime_{\rm CO}$ } 
\label{sec:alpha_co}

\begin{figure*}[htbp]
\centering
        \includegraphics[width=5in]{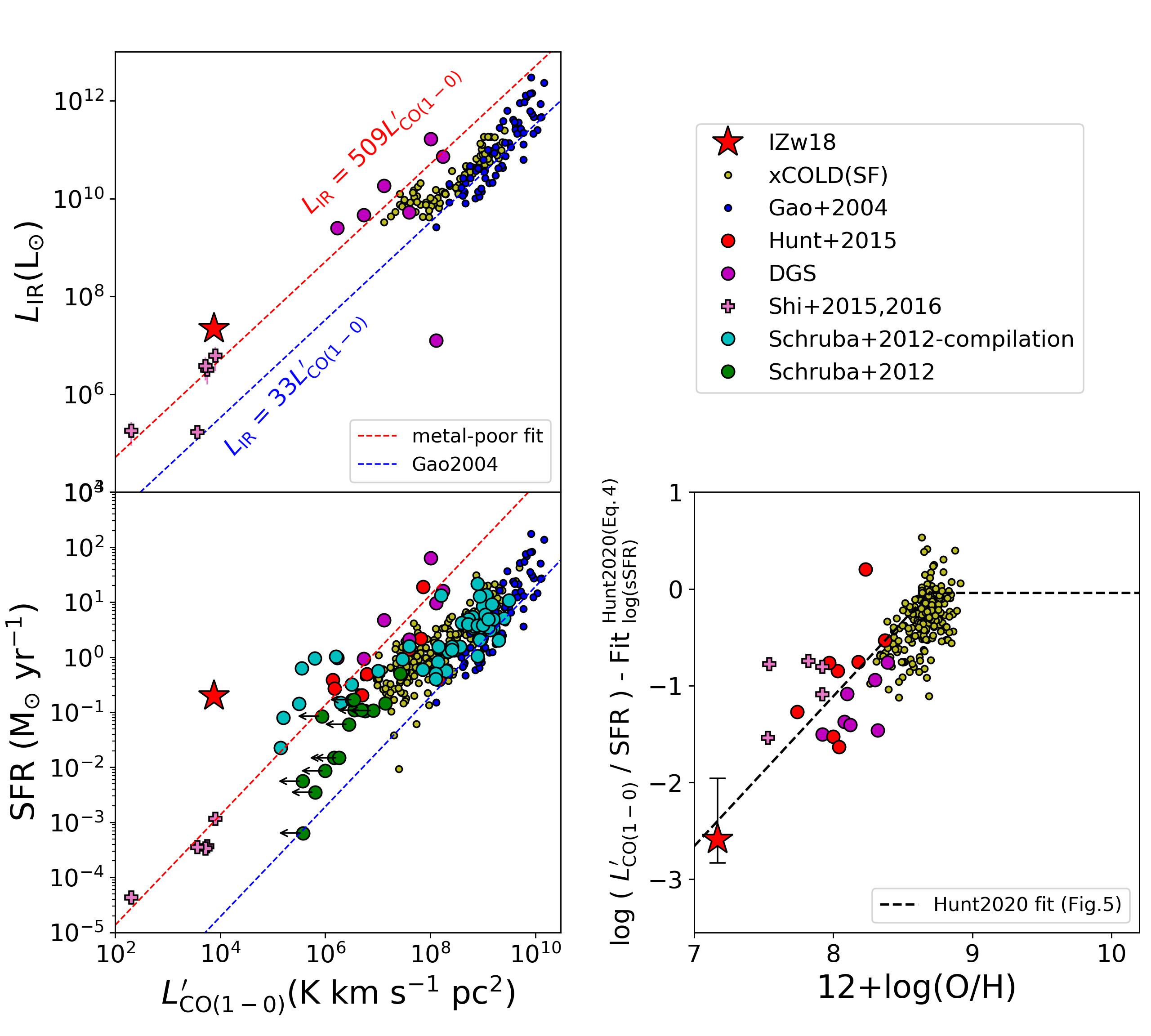} 
    \caption{Infrared luminosity (\textit{top left}) and SFR (derived from the emission at FUV and 24$\mu$m,  \textit{bottom left}) versus $L^\prime_{\rm CO(1-0)}$. The upper limit of IZw18 is shown as a red star, in comparison with the galaxies described in Section~\ref{sec:discussion}. In the two panels on the left, 
   the blue dashed lines show the  correlation by \citet{Gao2004} and the red dashed lines show the linear fit of the metal-poor galaxies \citep[][DGS]{Shi2015, Shi2016, Hunt2015, Cormier2014}.  The bottom right panel shows the empirically derived broken power-law regression for $L^\prime_{\rm CO(1-0)}$/SFR as a function of metallicity, illustrating the decrease of $\alpha_{\rm CO}$  with metallicity as introduced by  \citet{Hunt2020}.  The error bar of IZw18 represents the uncertainty in the measurements of the stellar mass. We note that in the panels, all the circles represent the global properties of the galaxies, while the magenta crosses denote the star-forming regions.}
\label{fig:LirSFR}
\end{figure*}

In Fig.~\ref{fig:LirSFR}, we have plotted the total infrared luminosity (8\,--\,1000\,$\mu$m), $L_{\rm IR}$, versus $L^\prime_{\rm CO(1-0)}$ and the SFR versus $L^\prime_{\rm CO(1-0)}$.  These two correlations have been well established in massive star-forming galaxies \citep[e.g.][]{Gao2004} as  CO molecules and $L_{\rm IR}$ trace H$_2$ molecular clouds and dust emission, respectively.
We assume optically thick and thermalised CO emission, then $L^\prime_{\rm CO(1-0)}$\,=\,$L^\prime_{\rm CO(2-1)}$. We note that CO tends to become warm and optically thin as CO is exposed to interstellar radiation in metal-poor galaxies. \citet{Schruba2012} adopted $R_{21}$\,=\,0.7, which is the average value of HERACLES galaxies \citep{Rosolowsky2015} for nearby dwarf galaxies. \citet{Saintonge2011} studied the 523 galaxies in xCOLD GASS survey and gave an average value of 0.79$\pm$0.03 with a larger scatter at $L^\prime_{\rm CO(1-0)}$\,<\,10$^8$\,K\,km\,s$^{-1}$\,pc$^2$.  Therefore, the choice of $R_{21}$ may lead to the derived $L^\prime_{\rm CO(1-0)}$ a factor of 1.4 different from the one based on our assumption.  The SFR was derived from the combination of FUV and 24\,$\mu$m emission for all galaxies,  except for the infrared bright galaxies for which the SFR was derived from  infrared luminosity.

Compared to  star-forming galaxies with Solar metallicity, metal-poor galaxies show higher $L_{\rm IR}$/$L^\prime_{\rm CO(1-0)}$ and SFR/$L^\prime_{\rm CO(1-0)}$ ratios. The  $L_{\rm IR}$/$L^\prime_{\rm CO(1-0)}$ ratio 
of  metal-poor galaxies  is, on average, 15.4 times higher than the one of the  massive, normal star-forming galaxies at Solar luminosity.  IZw18  shows even higher    $L_{\rm IR}$/$L^\prime_{\rm CO(1-0)}$ and SFR/$L^\prime_{\rm CO(1-0)}$ ratios, which are $\gtrsim$\,5.7 and $\gtrsim$\,119 times higher than those of other metal-poor galaxies (Fig.~\ref{fig:LirSFR}-\textit{Top left}), respectively.  It is difficult for CO molecules  to form in metal-poor environments because there is fewer raw material and little dust to protect the already scarce CO molecules from  UV radiation. The higher ratios are a natural result of this. 
The SFR/$L^\prime_{\rm CO(1-0)}$ ratio of metal-poor galaxies increases as the metallicity decreases (Fig.~\ref{fig:LirSFR}-\textit{Bottom right}). \citet{Hunt2015} found a significant correlation when fitting their  dwarf galaxies,  detections in \citet{Schruba2012}, and the compilation therein. We see a slightly shallower trend when including all galaxies with a CO detection. We kept only the star-forming galaxies in xCOLD in the regression to exclude the contamination from active galactic nucleis.  The SFR/$L^\prime_{\rm CO(1-0)}$ ratio  of IZw18 falls above the trend  and it is more than ten times higher than the rest of the galaxies. This suggests a great difference in the CO distribution and content of  galaxies similar to IZw18, at a few percentages of solar metallicity, than that of metal-rich galaxies. However, as we discuss below, the change may occur gradually with the decreased metallicity. 

SFR and $L^\prime_{\rm CO(1-0)}$ can be  derived from observations, but the molecular gas mass, $M_{\rm H_2}$\,=\,$\alpha_{\rm CO}$\,$\times$\,$L^\prime_{\rm CO(1-0)}$, is the key quantity of interest. The observed  SFR/$L^\prime_{\rm CO(1-0)}$ ratio is related to both the depletion time ($\tau_{\rm dep}$) and the CO-to-H$_2$ conversion factor ($\alpha_{\rm CO}$) as SFR/$L^\prime_{\rm CO(1-0)}$\,=\,SFR/($M_{\rm H_2}$/$\alpha_{\rm CO}$)\,=\,$\alpha_{\rm CO}$/$\tau_{\rm dep}$. Many studies have shown that $\alpha_{\rm CO}$ increases rapidly at low metallicity  theoretically and observationally \citep{Israel1997,Glover2011, Leroy2011, Narayanan2012, Elmegreen2013, Shi2016}. If $\tau_{\rm dep}$ is constant, as is found for the nearby disc galaxies \citep{Leroy2008, Bigiel2011}, then the dependence of the SFR/$L^\prime_{\rm CO(1-0)}$ ratio on metallicity is the direct consequence of the variation of $\alpha_{\rm CO}$. However, $\tau_{\rm dep}$ in dwarf galaxies remains uncertain and tends to be shorter at a lower mass and at a higher specific SFR  \citep[sSFR;][]{Saintonge2011, Shi2014, Hunt2020}. IZw18, as a  dwarf galaxy with a relatively high sSFR 
and low mass,  is likely to have low $\tau_{\rm dep}$. 
We speculates that as $\alpha_{\rm CO}$ = $\tau_{\rm dep}$\,$\times$\,$\frac{\rm SFR}{L^\prime_{\rm CO(1-0)}}$, the high SFR/$L^\prime_{\rm CO(1-0)}$ ratio of IZw18 would overwhelm the possibly low $\tau_{\rm dep}$, and this would result in a high $\alpha_{\rm CO}$.
Moreover, considering the dependence of $\alpha_{\rm CO}$ on both sSFR and metallicity for the metal poor galaxies, IZw18 follows the empirical relation found in \citet{Hunt2020} within the uncertainty of the stellar mass, as shown in the bottom right corner of  Fig.~\ref{fig:Lcii}. This relation was derived based on a recent compilation of $\sim$400 metal poor galaxies \citep[][MAGMA]{Ginolfi2020}. We note that the individual star-forming regions of the metal-poor galaxies from \citet{Shi2015, Shi2016}  all fall slightly above the relation, but well within the scatter of the global measurements of the  galaxies. This indicates that $\alpha_{\rm CO}$ changes continuously with metallicity and sSFR. 
A further constraint on the conversion factor ($\alpha_{\rm CO}$) of IZw18 is beyond the scope of this paper. 

\subsection{The structure of the interstellar medium}
\begin{figure}[htbp]
\centering
        \includegraphics[width=\columnwidth]{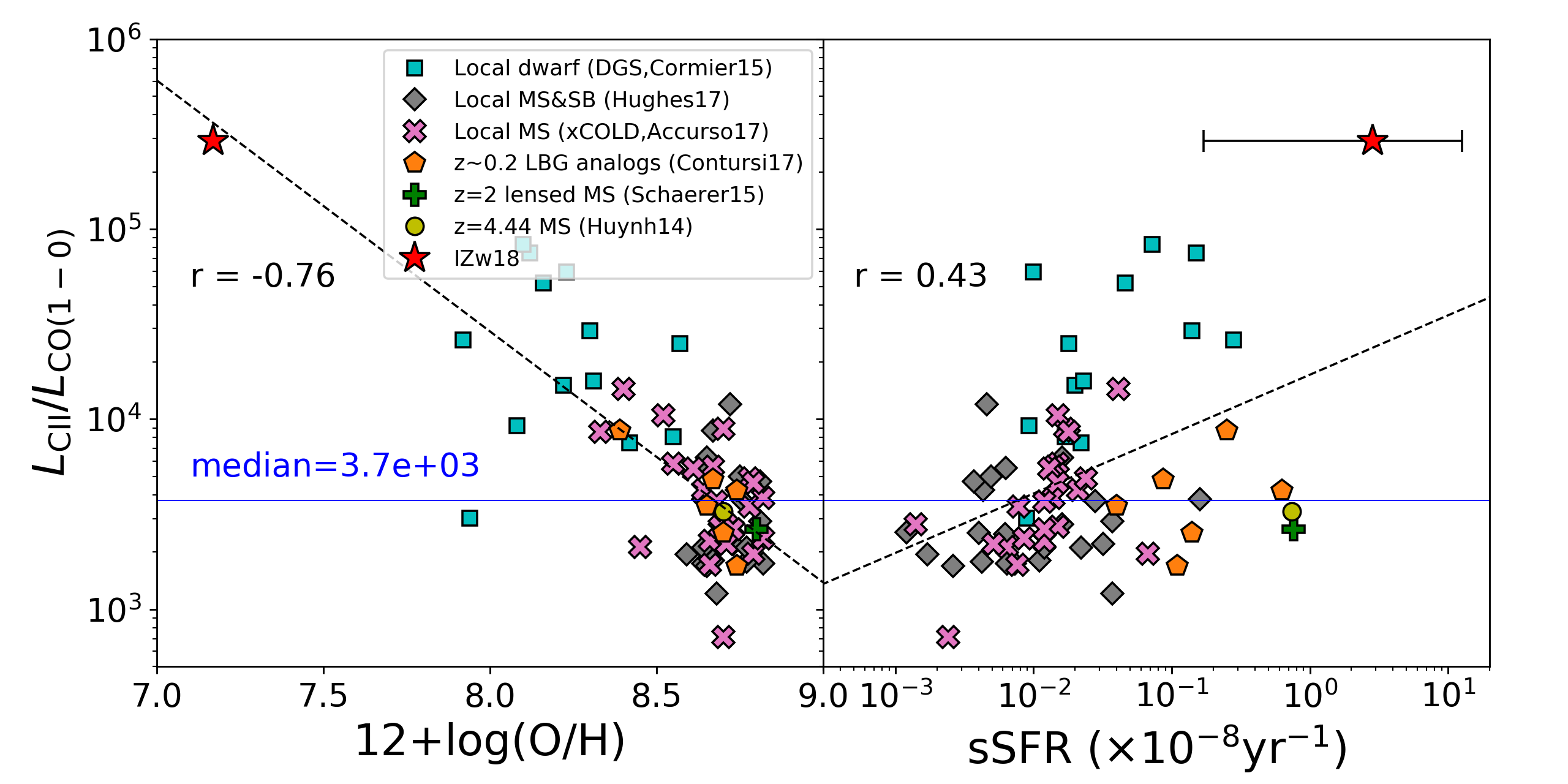} 
    \caption{ $L_{\rm CII}$/$L_{\rm CO(1-0)}$ ratio as a function of the metallicity (\textit{Left}) and specific SFR (\textit{Right}).  IZw18 (red star) is compared  to the galaxies compiled in \citet[][see references therein]{Zanella2018}. The errorbar reflects the uncertainties of stellar mass and SFR.   Pearson's correlation coefficients for the regression fits are shown in  each panel.} 
\label{fig:Lcii}
\end{figure}

\citet{Bolatto1999} and \citet{Rollig2006} modelled the PDRs and found that as the [C\,\textsc{ii}] regions expand at low metallicity,  the $L_{\rm [C\,\textsc{ii}]}$/$L_{\rm CO(1-0)}$ ratio increases. This has also been confirmed in previous observations \citep{Madden2000, Cormier2014}.  \citet{Cormier2015} detected the bright [C\,\textsc{ii}] emission in IZw18. We plotted the $L_{\rm [C\,\textsc{ii}]}$/$L_{\rm CO(1-0)}$ ratio as a function of metallicity and specific SFR in Fig.~\ref{fig:Lcii}. IZw18 generally follows the trend with metallicity, defined by the galaxies compiled by \citet[][see references therein]{Zanella2018}, but it shows ratios several times higher than the prediction of the regression fit to sSFR. We note that 
[C\,\textsc{ii}] is considered  to be a good tracer of CO-dark molecular gas \citep[e.g.][]{Cormier2015, Accurso2017, Zanella2018, Madden2020}. Even though [C\,\textsc{ii}] can originate from both ionised gas and neutral gas, simulations and models have found that the ionised fraction does not go beyond 50\%, even for metal-poor galaxies \citep{Accurso2016, Cormier2019}. The ionised fraction  of IZw18 is unknown, but it could be low as recent studies  found a decreasing fraction with decreasing metallicity \citep[][and references therein]{Madden2020}. 
This indicates that the total molecular reservoir, traced by [C\,\textsc{ii}] emission from PDRs, is much larger than the CO emitter in IZw18. 
Then the extreme case of IZw18 reinforces a significant change in the ISM structure  in systems at a few percent of the Solar metallicity and this may have similar implications for such systems in the early Universe.

\section{Conclusions}
\label{sec:conclusion}
In this letter, we report a marginal detection of CO $J$=2-1 in IZw18, using the observation from NOEMA  after its Phase II upgrade. We pushed  the detection limit of $L^\prime_{\rm CO}$  down to  $L^\prime_{\rm CO(2-1)}$\,=\,3.99\,$\times$\,10$^3$\,K\,km\,s$^{-1}$\,pc$^{-2}$,  which is 40 times lower than that of \citet{Leroy2007}. As one of the most metal-poor galaxies, IZw18 shows $L_{\rm IR}$/$L^\prime_{\rm CO(1-0)}$ and SFR/$L^\prime_{\rm CO(1-0)}$ ratios much higher than those of galaxies with a higher metal abundance. Particularly, the SFR/$L^\prime_{\rm CO(1-0)}$ ratio of IZw18 constrains the CO-to-H$_2$ conversion factor to rise considerably at metallicity lower than 5\%\,Z$_\odot$. The SFR/$L^\prime_{\rm CO(1-0)}$ ratio also follows the regression found in \citet{Hunt2020} well which considers the influence of both sSFR and metallicity.
The high $L_{\rm [C\,\textsc{ii}]}$/$L_{\rm CO(1-0)}$ ratio indicates that the CO emitter may trace only the inner part of the entire molecular gas  reservoir due to the extremely low metallicity of IZw18.
An upper limit of the continuum emission at 1.3\,mm is also obtained to constrain the Rayleigh Jeans tail of the SED and we excluded a sub-millimetre excess in IZw18. 


\begin{acknowledgements}
We are grateful to the referee for useful comments which significantly improved the quality of this manuscript.
L.Z.  and Y.S. acknowledge the support from the National Key R\&D Program of China (No. 2018YFA0404502, No. 2017YFA0402704) and the National Natural Science Foundation of China (NSFC grants 11825302, 11733002 and 11773013) and the science research grants from the China Manned Space Project with NO. CMS-CSST-2021-B02. Y.S. thanks the support by the Tencent Foundation through the XPLORER PRIZE. L.Z.  also acknowledges China Scholarship Council (CSC). J.W. is thankful for the support of NSFC (grant 11590783). 
\end{acknowledgements}

\bibliographystyle{aa} 
\bibliography{bibitem.bib} 
%

\appendix
\section{Multiwavelength images}
\label{sec:append}
\begin{figure}[htbp]
\centering
        \includegraphics[width=2\columnwidth]{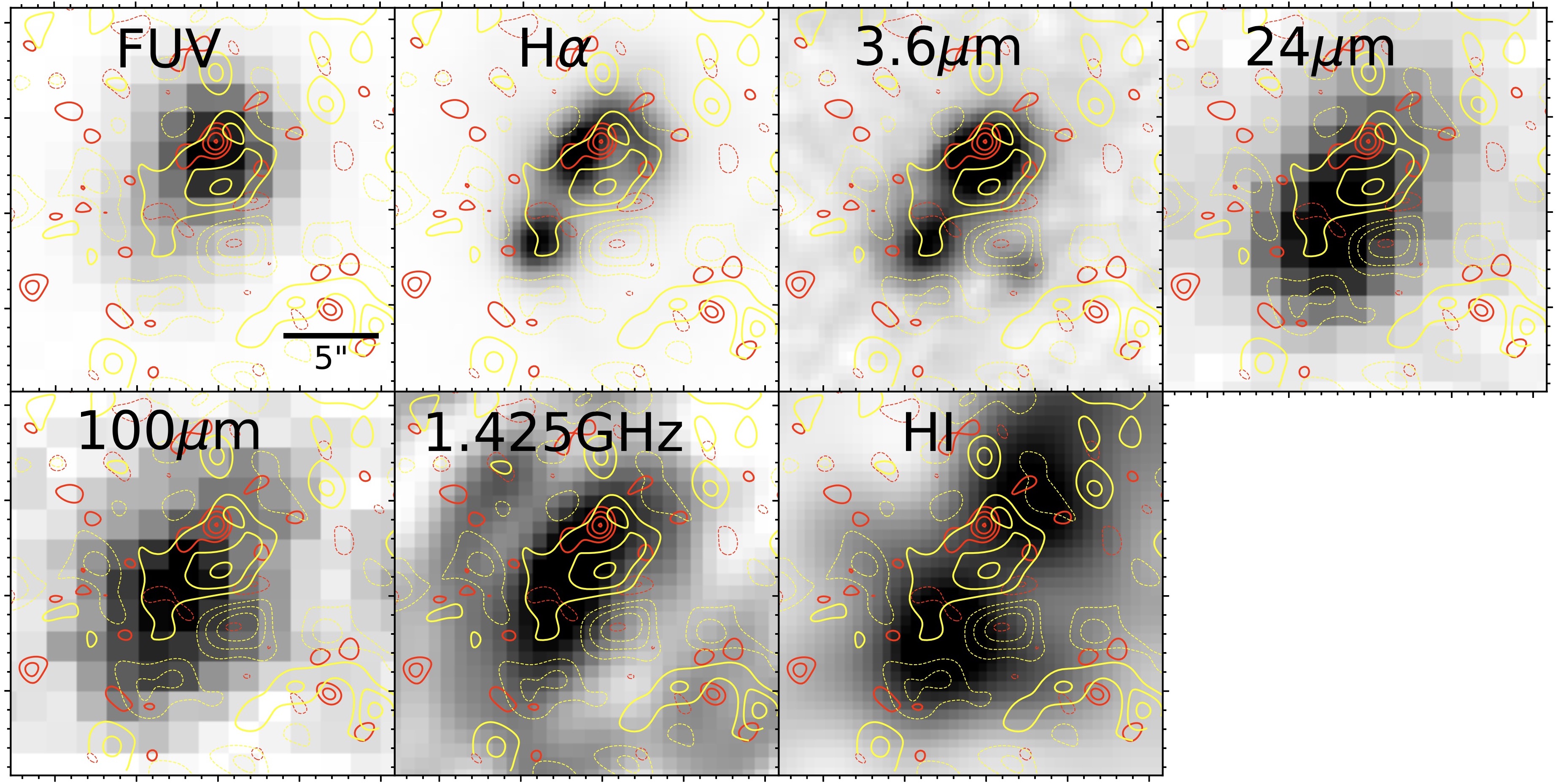} 
    \caption{Similar to Fig.~\ref{fig:imgV}, but CO $J$=2-1 emission and 1.3\,mm continuum superimposed on the $GALEX$ FUV \citep{GildePaz2007}, H$\alpha$ \citep{GildePaz2003}, \textit{Spitzer} IRAC1 at \,3.6\,$\mu$m \citep{Brown2014}, \textit{Spitzer} MIPS\,24\,$\mu$m \citep{Brown2014}, \textit{Herschel} MIPS at 100$\mu$m \citep{Fisher2014}, and VLA L-band at 1.425GHz \citep{Hunt2005} continuum maps, as well as the H\,\textsc{i} gas intensity map \citep{Lelli2012}. The CO emission coincides with  emissions at FUV, H$\alpha$, and radio continuum, which trace star formation, along with  stellar emission at 3.6\,$\mu$m. Meanwhile, the hot and cold dust emission at 24\,$\mu$m and 100\,$\mu$m show a slight offset ($\sim$3$\arcsec$) from the CO emission. The extended emission at 1.3\,mm generally covers the emission at all bands shown here, except for H\,\textsc{i} gas. Emission at the 1.3\,mm continuum falls between the two peaks of H\,\textsc{i} gas emission, and CO emission is also shifted from one of the H\,\textsc{i} peaks by $\sim$2\arcsec. A clear offset is shown between H\,\textsc{i} gas and CO gas and the 1.3\,mm continuum. An offset between CO gas and H\,\textsc{i} gas has also been observed in another metal-poor galaxy, Sextans~B \citep{Shi2016}. }
\label{fig:img}
\end{figure}

\end{CJK*}
\end{document}